# Energy Efficient Service Distribution in Internet of Things

**Barzan Yosuf, Mohamed Musa, Taisir Elgorashi, Ahmed Q. Lawey, and J. M. H. Elmirghani**
*School of Electronic and Electrical Engineering, University of Leeds, Leeds, LS2 9JT, United Kingdom
E-mail: {elbay@leeds.ac.uk, M.Musa@leeds.ac.uk, T.E.H.Elgorashi@leeds.ac.uk, A.Q.Lawey@leeds.ac.uk,
J.M.H.Elmirghani@leeds.ac.uk}*

**ABSTRACT**
The Internet of Things (IoT) networks are expected to involve myriad of devices, ranging from simple sensors to powerful single board computers and smart phones. The great advancement in computational power of embedded technologies have enabled the integration of these devices into the IoT network, allowing for cloud functionalities to be extended near to the source of data. In this paper we study a multi-layer distributed IoT architecture supported by fog and cloud. We optimize the placement of the IoT services in this architecture so that the total power consumption is minimized. Our results show that, introducing local computation at the IoT layer can bring up to 90% power savings compared with general purpose servers in a central cloud.
**Keywords**: IoT, MILP, service distribution, fog, cloud, edge.

## 1. INTRODUCTION

The Internet of Things (IoT) has become an emerging paradigm, in which billions of every-day objects ranging from simple sensors, actuators and smart devices are expected to be connected to the Internet. Cisco projected that by the year 2020, the number of connected IoT devices would reach around 50 billion [1]. These connected devices will be used in various applications such as but not limited to smart surveillance, e-health, smart city, smart homes [2]. In the past, these IoT type devices usually included sensors and actuators that were equipped with very simple CPUs and transceivers. However, due to the advancement in embedded technologies, devices such as smart phones and single-board computers like the Raspberry Pi have become more intelligent; hence, a substantial increase is observed in their communication and computational power consumption [3]. The increase in the intelligence of these smart devices implies that, IoT service requests can be fulfilled locally using on-board resources. However, IoT device resources still face limitations due to stringent battery lifetime, CPU and storage capacity limitations [4]. These limitations impede hosting resource-intensive tasks and storing large amounts of data locally on the IoT devices. Thus; cloud computing is utilized to cater for limitations in the IoT layer by executing resource intensive tasks and storing large amounts of data [5]. A typical cloud is concentrated in the core network and its resources are accessed through a transport network [6].

Fog computing, is a new model in which edge devices closer to the source of data are utilised to extend the functionality of the cloud [7]. It is worth mentioning that, the fog is not an alternative to the cloud but rather these two paradigms complement one another. The complementary features of the cloud and the fog, pave the way for a new breed of next generation IoT applications and services [1]. Existing studies mainly focus on the optimization of individual layers of the IoT network i.e. the end-device layer, access layer or the core layer separately. While it is important to focus on the mentioned layers individually and understand how to maximize their efficiencies, we believe that; it is of greater importance to focus on a bigger picture by optimizing the layers of the entire infrastructure concurrently [8]. The work in [2] formulates the service distribution problem in an IoT-Cloud framework using linear programming. The authors define the problem whose solution results in the optimum placement of IoT service functions and the routing of network flows across a heirachical architecure that satisfy end-user demands while minimizing energy consumption. Our previous work presented energy efficient solutions in cloud and core networks using mixed integer linear programming techniques considering a variey of scenarios including renewable energy sources, topology design, core networks with data centres, content distribution, network coding, and virtualization [9]-[18]. The remainder of this paper is organized as follows: In Section 2, the multi-layer IoT architecture is described. Section 3 presents results and discussions followed by Section 4 in which we provide our consulsions.

## 2. Multi-Layer IoT Architecture

Figure 1 shows the end-to-end network connecting IoT devices to the central cloud. The network consists of three layers: the access network, metro network and core network. We consider a multi-layer IoT architecture supported by fog and cloud. The architecture is composed of four main layers. The leftmost layer of the architecture comprises of the IoT devices. These devices are smart, wireless nodes that are used to collect data and transmit the same to the next layer(s) for processing and/or storage. The Access Fog layer comprises of wireless access points that are used to aggregate service requests from the IoT, either to be forwarded to the next layer(s) or to be hosted locally for processing. The Metro & Edge Fog layer contains a micro-data center located at the OLT which can be used to process the aggregated IoT service requests [8]. The rightmost layer is the Cloud layer with high capacity servers in a data centre.

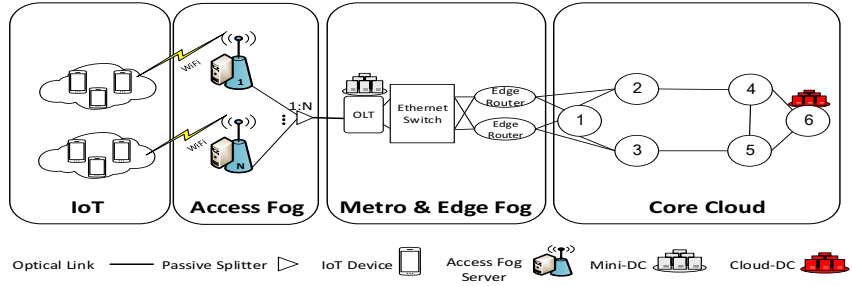

*Figure 1: A multi-layer IoT architecture, supported by fog and cloud platforms.*

### 3. RESULTS AND DISCUSSIONS

As mentioned previously, we consider a multi-layer architecture between the IoT devices and the Core Cloud. The IoT devices are the sources of service requests and we adopt the notion that a request per IoT device cannot be split between the computing layers. However, we allow for the IoT devices to be distributed between multiple layers, i.e. $x$ proportion of IoTs may choose local computation while $y$ proportion chooses remote computation. A service request consists of demand for processing (in MIPS) and bandwidth to transmit data (in Mbps). We assume that all service requests are homogenous among the IoT devices. We consider 25 IoT devices, which are randomly and uniformly distributed among the access points and each IoT device is located between 10 – 50 meters away from its respective access point. The distance is used to calculate the power consumption in the wireless domain due to amplification requirements which are distance dependant [19]. In addition, we assume all IoT devices communicate through WiFi with a supported bit rate of 150 Mbps [20]. It is important to note that we only consider uplink traffic, as this is reported to carry a much larger proportion of the data generated by IoT networks [19]. Furthermore, we assume the number of CPU instructions required to process one bit of data remains constant throughout our evaluations. The data about processing requirements was extracted from the work in [2] where it is reported that a typical analysis and decision making application in IoT requires 750 instructions to process one bit of data. For the traffic data, we consider a range from 0.5 – 4.5 Mbps, which is typical data for low – high quality videos in cctv streaming appications [21]. We consider the Rasbperri Pi Zero to represent the IoT devices, thus, we take its values for processing and networking as input parameters in our model. We estimate the processing capability (in MIPS) of IoT devices based on the data extracted from the technical report in [22]. It is reported that high-end intel CPUs execute 4 instructions per clock cycle (Hz). We believe that it is reaonable to assume that IoT type CPUs execute a single instruction per clock cycle (Hz), hence, the number of instructions executed per Hz is multiplied by the device's clock frequency (GHz) to estimate the total CPU capacity in MIPS. We also use the power consumption data of the Raspberri Pi Zero to calculate the efficiencies (processing and networking). We consider two types of cloud servers, a general purpose server (GP) and a special purpose server (SP). The general purpose servers are designed to execute a range of generic services, hence; they are not optimized for a particular service. In contrast, a special purpose server is purpose built; e.g. highly optimized to undertake a given task at any instance of time which in turn provides a better performance. An example of such special purpose servers are the GPU based servers used for mathematical processing and analysis based requests, such as the one provided by Nvidia [26], which are reported to be 10 times more efficient than their general purpose counterparts. We developed a MILP model to optimize the placment of the IoT services in the multi-layer IoT architecture so that the total power consumption is minimized.

*Table 1. Input Parameters for Processing Devices.*

| Device | MIPS | Power Consumption | | Instruction/ Hz | Freq. GHz | Device Layer | Num. of Nodes |
|---|---|---|---|---|---|---|---|
| | | Max (W) | Idle (W) | | | | |
| RPi Zero | 1000 | 3.6 [23] | 0.45 [23] | 1 | 1 [20] | IoT | 25 |
| RPi 3 | 2400 | 12.5 [23] | 2 [23] | 2 | 1.2[24] | Access Fog | 8 |
| GP Server | 10800 | 363 [17] | 112 [17] | 4 | 2.7[17] | Edge Fog/ Cloud | 1 |
| SP Server | 108k | 363 [17] | 112 [17] | 40 | 2.7 [17] | Cloud | 1 |

*Table 2. Input Parameters for Networking Elements.*

| Device | Bitrate (Gbps) | Power Consumption (W) | | Device Layer |
|---|---|---|---|---|
| | | Max | Idle | |
| WiFi Interface | 0.15 [20] | 1.2 | 0.45 [23] | IoT |
| ONU + WiFi Interface | 0.3 [16] | 15 [25] | 9 | Access Fog |
| OLT (1-port) | 2.4 [7] | 48 [7] | 43 [7] | Edge Fog |
| Ethernet Switch | 320 [8] | 3.8k [8] | 3.42k | Metro, DCs |
| Edge Router | 560 [7] | 4.55k[7] | 4.01k[7] | Metro, DCs |
| IP/WDM Equipment | 40 | 1.15k | 1k | Core Network |

The access network is a GPON network comprising 8 ONU nodes. GPON is particularly energy efficient and is expected to be increasingly used [8]. Each ONU has a built-in WiFi interface that is used to communicate with the IoT layer and is constrained by the maximum bit rate of 300 Mbps [25]. The ONU devices share a single fiber optic line to the OLT node with a split ratio of 1:8. We assume backhaul capacity of ONUs is sufficient for the present study. We assume ONU devices have similar processing capability to a Raspberry Pi 3, which we estimate to process 2 instructions per clock frequency (Hz). Moreover, the total CPU capacity (in MIPS) per Access Fog device was estimated in a similar way to the IoT. Furthermore, an industrial IoT gateway device like the Dell Edge 3000 was also used to help us estimate the CPU capacity (in GHz) of the devices in the Access Fog layer. We assumed Access Fog devices employ a higher CPU capacity than the IoT devices, to account for the difference in efficiencies offered by these two layers.

We assume the micro-data center at the OLT consists of 10 high-end general purpose servers connected by a LAN network for aggregation purposes [7]. We adopt the data in [18] about the properties (power consumptions and CPU frequency) of a general purpose server. However, we still make use of the data provided in [22] to estimate the CPU capacity (in MIPS) of the general purpose severs. The total MIPS of these types of servers is assumed to be 10800 MIPS in total, hence, 4 instructions per clock frequency (Hz). The metro network consists of switches and edge routers, which aggregate incoming traffic from the OLT node en route to the cloud data center located at the core network.

The core network is considered to be an IP/WDM network and it consists of two layers, the IP layer and the optical layer. In the IP layer, an IP router is deployed at each node to aggregate access network traffic. The optical layer is used to interconnect the IP routers through optical switches and IP/WDM technologies [18]. For modelling simplicity, we have summed the power consumption of all of the core network components and used the energy per bit approach as described in [18]. We assume a centralized cloud data center is located at node 6 and the number of computing servers is unlimited. As mentioned earlier, the network within the centralized cloud is like that of the micro-data center. All the values used in the MILP model for processing and networking devices, are summerized in **Error! Reference source not found.** and *Table 2*; respectivly.

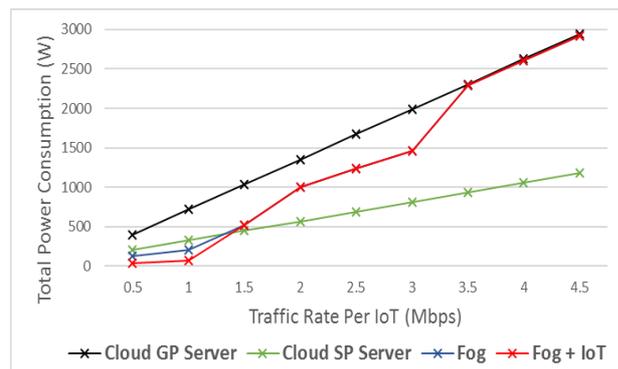
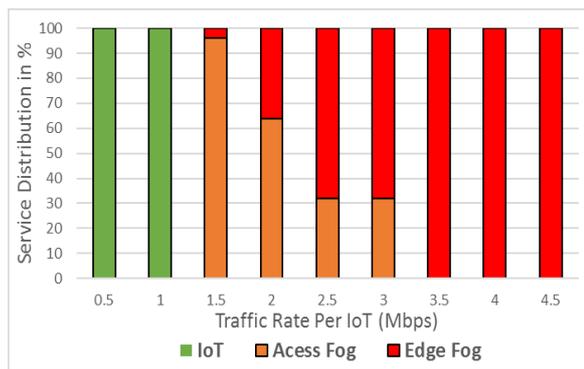

*Figure 2: Total power consumption*    *Figure 3: Distribution of service demands.*

Figure 2 shows the total power consumption of 4 scenarios considered in our service distribution model. The power consumption profile of networking and processing devices consist of two parts, an idle part and a load proportional part. We adopt the notion of energy per bit (traffic) and energy per instruction (processing) to calculate the power consumptions of the load proportional power consumption. We also assume that the idle power of low capacity network components such as the ONUs used in the Acess Fog layer is 60% of the maximum power consumption and likewise, 90% for high capcity equipemts such as OLTs and Ethernet Switches [7]. Scenario 1 (Cloud GP server) is the baseline approach and is the case where all IoT service requests are directed towards the conventional cloud data center located in the core network. Scenario 2 (Cloud SP Server) is simply opting for special purpose servers in the centralized cloud. The results show that a power saving of up to 59% can be achieved and 57% on average compared to processing services using general purpose servers. However, this superiority may come at a relatively higher delay cost compared to a distributed processing infrastructure (scenarios 3 and 4).

In scenario 3 (Fog), we introduce the concept of fog computing to the model and observe the efficiency of this approach. We assume in this case that IoT devices cannot be utilized for processing and thus only the two fog layers can be used. It is shown that for data less than 1.5 Mbps, up to 67% power saving can be obtained. This is due to the efficiency of the devices used in the Access Fog layer, however this saving diminishes as data traffic increases due to the capacity constraints of processing at the Access Fog layer. Hence, once the Access Fog layer capacity is exceeded, IoT services are transmitted towards the micro-data center which employs general purpose servers at the OLT node and the average saving is 18%. Scenario 2 (Fog + IoT) considers local computation at the IoT devices in addition to the fog. As shown in Figure 2, for low data rates, processing on IoT devices can yield a power saving of up to 90%. This is due to the fact that services are kept local and the heavy cost of the transport network is avoided. On average, this approach yields 19% savings due to the limited processing capacity of local computation.

Figure 3 shows the placement of the IoT services of different data rates. As the processing demand is proportional to the traffic demand, IoT services of higher data rates are processed in higher layers of higher processing capacities.

## 4. CONCLUSIONS

In this paper, we studied the problem of service distribution in a multi-layer IoT architecture. We minimized the total power consumption by optimizing the distribution of IoT services between several layers of processing. For services with low computing demands, savings can be as high as 90% if all tasks are processed by IoT devices compared to cloud with general purpose servers. Furthermore, for service demands that violate the processing capacity of IoT devices, Access Fog processing was introduced and this achieved a maximum power saving of 67%.

## ACKNOWLEDGEMENTS

The authors would like to acknowledge funding from the Engineering and Physical Sciences Research Council (EPSRC), through INTERNET (EP/H040536/1) and STAR (EP/K016873/1) projects. The first author would like to acknowledge his PhD funding from the Engineering and Physical Sciences Research Council (EPSRC). All data are provided in full in the results section of this paper.